# Applying whole system design in a sportscar factory

*Luca Piancastelli, Leonardo Frizziero and Simone Marcoppido,*
DIEM, University of Bologna, viale Risorgimento, 2 - 40136 Bologna, Italy, e-mail: luca.piancastelli@unibo.it

*Eugenio Pezzuti,* University of Rome Tor Vergata, Faculty of Engineering, via del Politecnico, 1 – 00133 Rome, Italy,
e-mail: pezzuti@mec.uniroma2.it

*Basing on a paper which explores the adoption of a whole system approach to a more sustainable and innovative design, the present paper wants to apply the same approach to a real case, inside of a famous Italian sportscar factory. A case study in this factory was developed and decodified gaining improved understanding of whole system design and those factors that substantially influence its success. All the factors mentioned above (such as dynamics of flattened hierarchy, the need to identify relationship between parts of the system) are used, into the application presented in this paper, to achieve an ultimate optimization of the whole.*

The world crisis had accelerated the necessity to make innovation; innovation must be realized into new products; new innovative products can be thought only new deisgn process. We can say also that the growing issue linked to increasingly complex problems, combined with concerns for the environment, is fuelling the demand for more innovative and sustainable products, services and systems. Designers are used to adopt more holistic and integrated approaches in an attempt to meet increasing consumer demands (Coley & Lemon, 2008). An innovative approach to the development of an optimised solution requires to be trans-disciplinary. It could be very valuable if different kind of phenomena (social, economic and enviromental) will be analyzed and considered in order to develop an innovative and global design solution. The trans-disciplinary approach necessarily requires the cooperation of people with different backgrounds, coming from different sectors. The establishment of partnerships among all these actors will lead to an optimized whole system solution. These are the principles on which the process of "whole system design" is based (Stasinopoulos, Smith, Hargroves, & Desha, 2009). However, there is limited research concerning the integrative process that actors are required to follow in order to reach such a solution.

Many of the cooperation techniques, developed during the past two decades, have been recognized as key factors into the application of the new system level design approach. In fact system level design approach (as well as collaborative and sustainable ones) often requires, as mentioned above, the development of partnerships (Katzenback & Smith, 1993), the use of trans-disciplinary skills (Gibson, 2001; Postrel, 2002), and systems thinking approach (Senge, 2006). All these techniques, and their theorical implementation, have been interesting subjects for a wide range of different papers and publications.  However only little papers have been written on how these techniques are efficiently implemented, within a whole system design approach, into an industrial real complex design process. This paper aims to explain the implementation of the whole system design approach, for the development of sustainable and innovative solutions to complex design problems. For this purpose it will be presented a case of study based on a famous Italian sportscar factory, which commonly uses a whole system design approach to develop its innovative products, in order to conduct a detailed and indepth analysis of the implementation of this design process. Through the use of multiple examples of design meetings, the analysis of project documentation and interviews and discussion with project members, a number of factors were observed which appeared to be common to whole system design (Charnley & Lennon, 2011).

These factors were then confirmed, modified and validated through use of a number of anecdotes across multiple design contexts in the present sportscar factory. These experiences and anecdotes are presented and utilised to demonstrate the factors necessary to facilitate good whole system design. The study wants to integrate the study of Charnley & Lennon, 2011, providing an example of  application of the whole system design in a famous Italian factory of sportscars.

## *1 State of the art*

It is under common attention that contemporary society is profoundly changed; this change has led to more complex problems, incorporating multiple interconnected aspects, such as social, economic and environmental ones. Consequently, there is an increasing responsibility to replace incremental improvements to existing products with all-encompassing, sustainable and innovative packages of products, services and systems that will provide solutions to consumer needs and requirements (Bhamra & Evans, 1997; Brezet, 1997; Lofthouse, 2004).

Conventional businesses are launching new green initiatives and eco-friendly products each week in an effort to capitalize on society's apparent shift toward a more environmental ethic. However, there is a concern that by focusing on environmental sustainability alone, considerable opportunities for improved efficiency, innovation and functionality are being missed. Nowadays the environmental impact have become very important and one of the key characteristic of the new products. However these environmental characteristic of the products are still considered as an add-on to the product, an easy way to promote it as innovative and eco-friendly, and not the central theme of the whole design process, or at least one of the important aspects that has to be considered (Stasinopoulos et al., 2009). This consideration leads to a key statement which need to be considered: the design process we will use to develop a new product could have a deep environmental, social and economic impact. This idea should be the base for the understanding of sustainable development, and to achieve more awareness of its more important aspects (Howarth & Hadfield, 2006). The whole system design process should be explored , now more than ever, in order to develop a new design process that could really take in account of all those aspects, that will be critical in the next future. However most of the whole system design approaches used to develop a new "green" product still lead to a simple modification of traditional products, for example focusing on reducing its present environmental impact, rather than to the development of a new sustainable production process of the product (Morson, 2007).

The inability of analyzing the world as a whole, and so to understand deeply how the business is deeply linked to social, economical, environmental aspects, which are also linked one to the other, led to the modern un-healtiness Senge (2006). Sustainable development has been proclaimed as one of the most important objectives to be reached and to work for, but everyone tends to limit its efforts to what can be done within the boundary of its reality (Ehrenfeld, 2003). The boundary limited approach of the modern efforts will hardly lead to the achievement of the sustainability, because it doesn't deeply influences the whole society, from which every worker arrives, carrying on the relative cultural structure. What is requested to the business is to contribute to the change of this modern cultural structure, this could be done changing the products to facilitate the shift of the cosumer values and thus its demand. (Morson, 2007).

Anarow et al. (2003) expressed the idea that sustainability could be reached only if the society moves towards a whole system thinking. But as we said before the whole system thinking is the base of the new design processes. So to achieve the development of sustainable solutions it is necessary to change the way we think about design, moving towards a whole system thinking approach. In order to reach this important objective, in order to be more competitive on the market, many companies started to establish various types of partnerships across different industrial sectors. However this approach isn't enough to achieve a whole system perspective, because the terms of their integration within a common holistic process, aimed to the development af an innovative sustainable solution, are often unclear. There are currently multiple terms being used to describe holistic and integrated approaches to the design of more radically innovative and sustainable solutions (Coley & Lemon, 2009).

Although they often adopt a slightly different focus, these approaches have many attributes in common. Whole system design is one such approach which is becoming increasingly popular, however, there is limited research detailing the process that actors are required to follow in order to reach a sustainable, innovative and 'system level' solution (Charnley & Lennon, 2011).

## *1.1 Definition of whole system design*

As suggested by the Rocky Mountain Institute (2006), we can affirm that whole system design can be defined as

*"Optimising not just parts but the entire system. it takes ingenuity, intuition and team work. Everything must be considered simultaneously and analysed to reveal mutually advantageous interactions (synergies) as well as undesirable ones"*

Indeed it is very rare that the issues we have to face are simple and involves only one industrial sector, more often it is requested to solve complex problems, characterized by more than one critical aspect. So it is common that the issues we have to solve involve economic, social and environmental issues. In order to find the best sustainable solution to this type of problems it is necessary to have a trans-disciplinary approach, because the need to focus on the entire system, according to the whole system design process, requires the views of experts from the wide group of disciplines involved. More often it isn't possible to find a simple solution and it is necessary to mix together different systems, products and services in order to achieve a sustainable optimized solution, if exists. However the process neede to get to that solution it is often unstructured and messy (Charnley & Lennon, 2011).

Incremental improvements to existing systems rarely meet rising consumer expectations for solutions to be both effective and environmentally benign (Blanchard & Fabrycky, 2006). It is therefore suggested that to tackle increasingly complex and disorganized problems; to provide holistic and integrative solutions, we need to adopt a change in design mentality and to start thinking differently (Hawken, Lovins, & Lovins, 1999).

However it won't be easy to achieve this important change of mentality, the reason of this difficulty is inherent to the way, designer and engineers, approach to the problems and the way they search the solution. As scientific and technological knowledge grew during these years, designers, engineers and managers became more specialised. With this increasing specialization they began to use a solving approach based on splitting up complex problems in may different sub-problems involving different disciplines, and focusing their attention on a specific sub-problem.

Designers and engineers followed highly structured and 'over the wall' approaches to design such as those prescribed by Forsberg and Mooz (1998) and Pahl and Beitz (1996). As a result engineers and designers are no longer trained across fields and thus no longer keep up with the latest breakthroughs in every field (Stasinopoulos et al., 2009). A separation of design functions and processes means that opportunities are often missed to optimise the whole system, which can lead to inefficient design, construction delays, oversized heating systems, higher costs and unnecessary environmental impacts (Anarow et al., 2003). Stasinopoulos et al. (2009) suggest that this is largely due to the fact that the engineer only knows their field in detail and has little interaction with other designers on the project.

Due to the need of a systematic approach of addressing complex problems, from different prerspective, determined a growth of the importance of trans-disciplinary collaborations and partenerships, especially withinh the industry (Hebel, 2007; Senge, 2006). However, partnerships are accompanied by numerous expectations and requirements, and also a more extensive network of actors. Some actors, who were never previously regarded as designers, are becoming heavily involved with the actual process of designing. High levels of multi-disciplinary working not only increase levels of complexity (Mankin, Cohen, & Fitzgerald, 2004) but also create many more issues and concerns to consider and often they can be conflicting (Howarth & Hadfield, 2006).

## *1.2 Definition of system*

The main issue about the systems is that, as Seiffert and Loch (2005) suggests, they are a complex structure composed by many deeply interlinked parts. If we consider the world as a system, it is easy to individuate many problems that are closely linked one to each other without sharing any simple local cause. These type of problems are what Senge (2006) calls systemic breakdowns, the most critical aspect of this situation is that the complexity of the problems continues growing, so there is an increasing need of a system approach in order to manage this growing complexity. Whole system design approach is based on considering the problem as a whole system, and not to concentrate on its single components. In this approach the problem itself is considered to be created by one or some of the system's parts, but by every part.

Systems are conceptual devices that we bound with a purpose; however once bounded they become real and we can explore, and influence, how they emerge through internal restructuring and their interactions with their environment. The environment, in systemic terms, is that which lies outside of the system boundary. It is the ability to acquire and utilise information about that environment that forms the basis for an adaptive, and thereby more sustainable system (Lemon 1999). Anarow et al. (2003) recognise that a whole system approach focuses on interactions between the elements of a system as a way to understand and change the system itself. Without this whole system perspective crucial impacts between components could be missed, therefore disrupting the system as a whole and overlooking opportunities for improved efficiency and environmental sustainability. For the purpose of this paper, systems are defined as a set of independent but interlinked phenomena, or as Sherwood (2002) defines them 'a community of connected entities' that we bound with a purpose (e.g. the design process). This connectedness means that systems have emergent properties and cannot be broken into their component parts; we must consider them as a whole and therefore need to develop mechanisms for doing so (Charnley & Lennon, 2011).

## *2 Whole system design: process and results*

Referring to section 1.1, we have already said that there is little literature on how the whole system design is practically implemented into the industrial reality, how it can generate a succesfull holistic process and which are its most important success factors.So it was necessary to undertake a largely exploratory and inductive methodology to gain an expansive insight into the process of whole system design.

Referring also to the paper of Charnley & Lennon, 2011, entitled "Exploring the process of whole system design", in which an analysis inside the automotive sector was produced, to discover and to decode the whole system design methodology, we cite as following the main results obtained by the analysis above.

The aims of the study above were to confirm, modify and validate the findings that were emerging from the whole system design method theory, to gain individual experiences of undertaking a whole system design from a variety of perspectives on the field and to gain critical feedback concerning the findings from the automotive sector from professionals across a variety of design disciplines.

Once data deriving from the study of Charnley & Lennon, 2011, had been collected and transcribed where appropriate, thematic analysis was used to identify, analyse and report patterns (themes) within the data, as prescribed by Braun and Clarke (2006) in their six step process. This technique was decided upon as it was appropriate to an inductive approach in which patterns and themes can be identified from different sources of raw data. Additionally, as the process being observed was complex; consisting of phenomena from multiple disciplines, a thematic approach enabled the data to be analysed without being simplified; allowing the underlying complexity to remain accessible.

The record taking, of the data resulting from the observation of design meetings, has experienced and important improvement as the researcher continued to observe the meetings and to take records. At the beginning, as the researcher has no experience about whole system design, the data recorded were messy and complex, according to the initial difficulty to undersrand which were the most important informations on how the design process should be carried out, as a result the data recorder were long and detailed.  Together with the growth of the researcher experience into whole system design it has been possible to recognize patterns within the data. This permitted to create specific groups of data, based on the assignment of themes and sub-themes.

A theme captures something important about the data in relation to the research question, and represents some level of patterned response or meaning within the data set (Braun & Clarke, 2006). Therefore, within the context of the study, a theme was defined as a set of behaviours, actions or thoughts that were displayed by those participants being observed and interviewed and were perceived by the researchers as significantly influencing the process of whole system design. Braun and Clarke (2006) suggest that ideally there will be a number of instances of the theme across the data set, however more instances do not necessarily mean the theme itself is more crucial. Researcher judgement was required to determine what a theme was; however, the data set was coded by more than one researcher to ensure validity and reliability (Krippendorff, 2004).

For the above reasons, it is important to underline that from the study of Charnley & Lennon, 2011, it was able to indicate the main factors which affect the success of whole system design methodology; they are:

   a. *Forming and sustaining a partnership*
   b. *Human and non-human interaction*
   c. *Individual characteristics*
   d. *Understanding of purpose and process*
   e. *Alignment of interests*
   f. *Sense making and system boundaries*
   g. *Facilitating whole system design*
   h. *Integration*

In particular, our paper will highlight how actually these factors are discoverable in a great famous factory which produces luxury sportscars in Italy. From theory to practice.

# 3 Factors influencing the success of a whole system design detected in a sportscar factory

The themes presented in the following sections (raised up from the study described above) are intended to provide a provisional framework to guide designers through the process of whole system design; they are not defining characteristics of whole system design but are factors that contribute to good design practise.

Further, to each of one of these factors, it is reported a description and an application of the case relative to the most important Italian sportscars factory.

## 3.1 Forming and sustaining a partnership in a sporscars factory

In the the most important Italian sportscars factory, the partnerships are the centrepiece of the company.

In fact, they always follow the evolution of the design project. Sometimes, the partners fill missing offices and functions.

The relationship between the company and its suppliers is very close, since the link is the design project. At the end, the partners (i.e. suppliers) are awarded just like the employees (during an event named *Suppliers Podium*).

The present company considers its own main partners (or suppliers) as an important resource, fundamental for itself.

Many of them are not discussed for many years. The relationship is solid. The "penetration" between them and the company is constant and continuous..

Some of the historical partners and suppliers of the present Italian spotscars company "garrison" continually the territory of the company, such as to be ready *just in time* for an immediate demand of the design project, paying the due to be considered ridiculous and lazybones, because they pass great part of their times immobile without doing anything along the corridors of the factory.

However, this bad image is immediatly rehabilitated when a problem (essential for the company) is solved by some of these parteners/suppliers.

For this reason is very difficult to become a supplier of this important Italian sportscars factory, because it is the same of becoming friend of somebody: you can become a friend of somebody, when he trust you; then, you can remain friend (quite) forever.

The study partecipants have noticed some important aspects which could be relevant for the correct and successful implementation of the whole system design in an industrial reality. As mentioned previously, the trans-disciplinary is one of the most important aspects of this design approach, so the development of partnership between different organization could be an important factor of success, because it facilitates the exchange of multiple different expert views and skills, and the recognition of the linkages between the system components. However all the partecipants have also highlighted that the development of a successful partnership it is not simple, as well as maintaining it.

In particular, it was observed that partnerships had been formed through the use of existing social and professional networks:

*"In my experience, the design team are known to each other, it is not always the same people but it is often the same companies involved in a project"* [Architect] (Charnley & Lennon, 2011).

Many of the partecipants have expressed the idea that a key factor for the establishment of a successful partnership could be the usage of previous existing relationship, both social or professional. Those previous relationship cold grant an important base of mutual trust and confidence, which represents, for most of the partecipants, one of the most important characteristic of a successful partnership. As opposed the development of a partnership with a new organisations, even if it could be exciting and potentially providing more informations, it has been described as difficult and time consuming.

Within the automotive case, several participants suggested that the design team consisted of some stakeholders who were not entirely suitable to the design context and had been chosen due to convenience as opposed to relevant expertise. This was said to cause inefficiency and slow progress whilst the partnership learned to make the best of the expertise available.

The development of valuable partnership, and its sustaining, has therefore been identified by the present paper as a key factor for a successful whole system design project. It has been observed that recruiting, and subsequently nurturing, the most appropriate experience and expertise for the design context can be overlooked or assumed, however, is necessary for a cohesive and successful whole system design team.

## 3.2 Human and non-human interaction in a sportscars factory

The management of the design project in the present sportscars factory happens in a such way that the same project is managed by an interdisciplinary team.

The progress and the updating of the project is carried out through a series of meetings, developing differetn issues about the project, which happen periodically and repeatedly every week.

The cadence of the meetings can be compared with the functioning of a Swiss clock; never, for any usefulness needlessly, these meetings must be cancelled or postponed.

The aim of the meetings is not only the one to update everyone about the development of the design project, managed with concurrent engineering methods, but also and mostly to create a team spirit, through the continuos interaction among its members, creating a feeling of togetherness, making eam building and giving to all the team clear target sto be reached.

The production of ideas deriving from the continuos team matching among all the members creates a virtuous loop which is a power both for the design project innovation and for the solution of small/big problems linked to the implementation of the same project.

The communications between all the actors during the meeting, also before and after them, has been recognized by all the parteicpants as one of the key factors of success of the whole system design project. An efficient and proactive communication has been identified as an important vector for the integration and for an active cooperation during the project.

Team members suggested that many of the delays within the automotive factory were due to actors not communicating their design decisions early enough in the process. It was stressed that design decisions need to be communicated, however small, as they may have a significant impact upon other components and ultimately affect the whole system. Participants in three of the six cases suggested that other team members were often unaware of the high levels of interaction required of them, particularly if a component was perceived not to be influential to their part of the design (Charnley & Lennon, 2011).

*"I can't be going to all the meetings because a lot of the stuff isn't relevant"* [Designer]

*"Unless he, as an architect, perceives that his design can benefit from talking with the engineers, there's nothing in the contractual arrangements that existed. So unfortunately industry is set up to avoid any of this (interaction)"* [Environmental Consultant]

Even if the design process should concern about the global design aspects, according to the whole design system approach, it has been often recognized the importance of digressions on detailed design areas. However, these digressions has often inhibited the open discussion and the participation during the meetings.

Several participants suggested that encouraging discussions to return to a system level during design team meetings encourages all to participate and share ideas; additionally this is when linkages between sub-systems are most likely to be identified.

*"I think partners need to be reminded at the beginning of every meeting that discussions are to be kept at a whole system level"* [Designer]

A lack of communication was observed, by the researcher, to inhibit the progress of integration and occasionally design decisions, which had not been communicated to the rest of the team, resulted in substantial delay later on in the process. To prevent this within future whole system design projects, it is recommended that actors should be made aware of the requirements and expectations that a whole system design process demands, early on within the project (Charnley & Lennon, 2011).

## 3.3 Individual characteristics in a sportscars factory

The design team of the Italian sportscars factory is composed by eleven people, each one with his own individual and professional characteristics.

The member of the team are:

1) Team leader
2) Team planner
3) Integration manager
4) Engine designer
5) Body designer
6) Quality manger
7) Cost engineer
8) Suppliers manager
9) Prototype manager
10) Purveyor
11) Marketing product manager

In particular we can say that:

1) Team leader is the chief and responsible of all the design project; he must relate about the progress of the project to the general direction.
2) Team planner can be considere the right hand of the team planner; he is the one who must keep all the project in order. His responsibility are the times and the costs, which must be related to the team leader.
3) Integration manager is the responsible of all the technicians of the team (who are the engine designer, the body designer and the prototype manager).
4) Engine designer is the technician who must integrate the engine into the vehicle.
5) Body designer must design chassis and body parts.
6) Quality manager is the responsible of all the quality processes during design and prototyping.
7) Cost engineer must follow the estimation of the costs of the design project.
8) Suppliers manager is responsible of the contacts between design team and suppliers.
9) Prototype manager is a technician who follows the construction of the prototype in the workshop.

10) Purveyor must buy the parts needed by the prototype manger for realizing the prototype.
11) Marketing product manager is the one who "defends" the interests of the costumers, indicating what kind of characteristics they prefer.

As it is easy to understand seeing the different functions of the team members, the design team receives different contributions and inputs for the implementation of the design project. These different inputs enhance the good outcome of the design project and the achievement of the targets.

Many of the team members referred to the process of whole system design as being different from a traditional design process and therefore requiring different skills.

*"It is a completely different skill set ... you have to be able to view things from the outside of the object, you have to be able to look down on the object"* [Architect]

Due to the trans-disciplinary nature of the whole system design process, it is obvious that a specific design decision will have an impact even on a large series of sub-systems, in addition of the impact on the specific sub-system. The capability to appreciate these "collateral" impacts is very important for the actors of the process, because it can lead to a deeper understanding of the process, understanding alla the linkages between the subsystems. Many of the participants have agreed that skills suchs as an open mind to new skills, the willingness to learn across boundaries and the capability of a systemic thinking could improve the capability of appreciation of the design decisions impacts. Another important opinion ,which has been expressed, relates to the fact that designer who owns the previously mentioned skills, among with an important expertise in their specific field could contribute more than other to make the whole system design process easier and successful.

*"I am sure that the role or prestige of the specialist has reached its absolute zenith... there is an ever increasing role for polymaths and I think the day of the polymath is returning because in whole system design that is the core skill"* [Designer]

The experience made within this project highlighted that those who were familiar with the traditional design processes, had experiences more difficulties in utilising trans-disciplinary skills. So it could be noticed that actors who participate in a whole system design process should have balanced skills between discipline specific and trans-disciplinary.

Additionally, skills such as thinking systemically are difficult to teach. It is suggested that sourcing actors who already possess and understand the benefits of utilising trans-disciplinary skills should be part of the recruitment process within a whole system design project. However, these skills are difficult to spot and therefore guidelines about how to identify those required characteristics should be developed early on. An example of this is the search for actors who display an enthusiasm to further their own learning and development and who show interest in areas aside from their own area of disciplinary expertise (Charnley & Lennon, 2011).

## 3.4 Understanding of purpose and process in a sportscars factory

As affirmed by Charnley & Lennon, 2011, also in the famous Italian sportscars factory findings early on in the study suggested that its employees found a whole system approach to be different to that of a more traditional design process, although it was implemented de facto. Furthermore, multiple participants of the study of Charnley & Lennon, 2011, within the automotive case in particular suggested that they were unsure of what a whole system approach was and what the benefits of undertaking the approach were.

*"I don't know what a whole system design is expected to be"* [Designer]

The same sentences are commonly said also in the Italian factory, object of the present paper. The designer's incomplete understanding of the whole system design approach, often represents one of the most important obstacles to the successful utilising of this design approach. It is true that it is not easy to explain this method to those who have never experienced it, because it requires a relevant grade of trans-disciplinary systemic way of thinking. However many of the Italian sportscar factory team members agreed if the whole system approach, that they were expected to adopt, had been comprehensively highlighted to them at the beginning of the project, along with the reasoning behind that decision, then it would have made the process easier to adopt. As a consequence of this considerations, the benefits of this design approach are often discovered later during the design process. So it would be important that those who had previous experiences of this design process, if there are some in the team, to focus on explaining the other members the benefit of this approach during the design process, in order to better highlight the purpose and the process of the project.

Team members said that it had been wrongly assumed that all of them were aware of what a whole system approach was.

One aspect that team members in Italian factory found challenging was the concept of emerging properties; i.e. qualitatively new situations that arise out of the interconnections within the design process. Parts of a whole system design often appear counter-intuitive unless the system is regarded as a whole.

*"Its completely counter-intuitive"* [Engine Designer]

It demonstrates the necessity to develop a shared sense of purpose and process within the context of the whole system, including its emergent properties. Often the benefits of a design cannot be seen as emergent properties are not being included; from this view it is counter-intuitive. Subsequently, the ability of team members to identify linkages between the components of a design, leading to the identification of emergent properties, largely influences the process of whole system design.

*13*
Applying whole system design in a sportscars factory

It was observed by the researcher also in the Italian sportscars factory that the more cohesive a team becomes the easier it is to form a shared understanding of purpose. However, it appears that the principles of whole system design are frequently misunderstood or unknown and therefore it should not be assumed that all actors have a shared understanding of the process required to reach a whole system solution.

## *3.5 Alignment of interests in a sportscars factory*

As remembered by Charnley & Lennon, 2011, a designer suggested that in traditional design the consumer and the manufacturer are at "polar opposites" in terms of needs and requirements.

Every one in automotive world (including the most famous Italian sportscars factory) usually says:

*"The manufacturer of a motor vehicle wants to make an ongoing profit through regularly servicing the vehicle once it is sold and relies on components needing to be replaced; to an extent failure is built into the design. On the other hand the consumer requires reliability, efficiency and quality and does not want to be frequently spending more money, time and effort on replacing components of the vehicle"*

Team members of the Italian company highlighted that establishing an alignment of interest between all parties involved in the process of a whole system design was "fundamental" and that "the opposition of interests is a terminal barrier to whole system design". During the project it has been recognized that the possibility to honestly talk about actor's interests and expectations on the paroject could greatly contribute to the successful achievement of a concrete, optimized and sustainable solution, capable of matching as many requirements as possible.

This is guaranteed by the Marketing Product Manager, who is a member of the team too.

The engineers and mostly the design project team members of the present Italian sportscars factory proposed that currently there is a lack of alignment between legislation and the aims of whole system design.

*"The vast majority of the industry will design to building regulations such as the Code for Sustainable Homes. But it should be right and it certainly can't stop the innovators from innovating which currently it actually, categorically, definitely, absolutely is doing and it should certainly not be pushing the noninnovators down a bad road which it is"* [Managing Director]

Not only an alignment of interests needs to be identified between a project team and its intended consumers, but alignment also needs to be sought from the policy makers that are imposing stringent targets and legislation upon those projects.

To achieve this alignment, it is recommended that requirements, needs, expectations and concerns of all partners should be discussed openly early on in the design process. Partners should be encouraged to be honest about what motivates and drives them. It was observed by the researcher that it is common at the start of a project for actors to 'keep the peace' by agreeing with shared goals; however this could lead to conflict later on in the process (Charnley & Lennon, 2011).

## *3.6 Sense making and system boundaries in a sportscars factory*

As the process of whole system design is frequently unclear it was assumed that sense making (Weick, 1995) would play a large part in that process. However, sense making activities often occur sub-consciously and it was therefore difficult for team members to recognise and relay specific accounts of when they had occurred (Charnley & Lennon, 2011).

In most of the case studies team members understood that it was necessary to spend time making sense of what a system was and where the system boundary should be drawn. Authors' suggested that often team members differ in what they perceive to be the most important aspect of a design. They recommended that to enable the team to make sense of the whole system, each team member should be asked to draw and detail the design from their perspective. This may enable a comprehensive architecture of the system to be developed which includes multiple perspectives and requirements. Additionally, based on the research, the authors' propose that the development of a shared architecture would enable actors to identify linkages between different components of the design more easily.

Also in the most famous Italian sportscars factory, the process of whole system design is complex due to the integration of multiple stakeholders and perspectives. Based on the findings of the research the authors propose that sense making techniques such as forming a common language and sense of purpose can assist with creating a project view and architecture. Additionally, defining a system boundary is a way of simplifying the complexity of a whole system design. However, it is recommended that this should not be used as an enduring structure as eventually the complexity of the system needs to be acknowledged.

Some of the mostly used expressions in the Italian factory are:

"*throw your heart over the obstacle*" [first user: Chief of Product Development]

"*criticalities of the project*" [first user: Team Leader 12 cylinders cars]

"*what have you got? the rava and the fava?*" [first user: Integration Manager]

"*do not we tell the story!*" [first user: Engine Designer]

## *3.7 Facilitating whole system design in a sportscars factory*

In the Italian sportscars factory, a substantial feeling of uncertainty can be felt amongst the design team, surrounding the process of whole system design, and design project team members usually suggest that this uncertainty is inhibiting design progress. Moreover, it is clear that uncertainty and ambiguity is the feeling diffused, mostly if related to the absence or a weak presence of a leader or manager.

*"So the house builders build their houses, the architects design them, the school workers deal with the school issues, the social workers deal with the youth issues but who on earth is supposed to manage the whole system? At the moment in East Leeds I have no idea who is overseeing that system approach"* (Charnley & Lennon, 2011).

The role of 'facilitator' was observed inside the most famous Italian sportscars factory. It is suggested that it is important, within a whole system design, to have an individual who is able to regard the system from above and to identify gaps or potentially overlooked relationships between sub-systems. Thi happens in the Italian factory through the presence and the role of the Team leader.

This is what happens in the Italian sportscars factory.

The role of the facilitator/team leader should not be confused with a leader or manager. It was observed that the most successful whole system design projects were managed through a flattened hierarchy. Participants from multiple cases agreed that a flattened hierarchy, consisting of stakeholders with equal shares in the project, encourages joint ownership and democratic governance. In the present Italian factory's teams it is clear how the encouragement of shared ownership amongst a design team leads to a feeling of empowerment and allows decisions to be made more efficiently. Additionally, members of a team were more likely to tie their identity to a project's outcome, thus injecting extra effort to ensure its success. This was observed throughout work and design inside the Italian sportscars factory and it is the authors' opinion that this goes a long way to supporting the process of whole system design. Cases in which a flattened hierarchy was successful also appeared to positively influence job satisfaction as team members said that they felt valued and their ideas were being recognised without having to get every single aspect signed off. The concept of shared responsibility and a flattened hierarchy was observed to work at its best when accompanied by a substantially integrated team.

## *3.8 Integration in a sportscars factory*

From the organization of the teams and of the design project in tha most famous Italian sportscars factory, it can be clear that the blurring of individual roles and disciplinary boundaries enables cross-disciplinary learning to be achieved and subsequently the impact of design decisions are more readily appreciated. However, it was identified early on in the automotive case that the blurring of roles can mean that responsibility is not accounted for.

*"Tasks are ignored and no one takes responsibility until eventually someone is forced to. Usually that task is not that person's role or responsibility"* [All the team members]

Subsequently this can result in components being missed out all together.

*"I always worry that we're missing something; that the organization is missing something. Obviously you can do the best you can but I always have this horrible feeling that there's going to be a gap between two bulk heads where a wire should be"* [All the team members]

*"You don't have to understand every single detail of how they work, its much more important to have a feel for what they do and how they fit into the system"* [Team Leader].

Moreover, in the most famous Italian sportscars factory it can be highlighted that for successful integration to take place, each actor needed to possess the skills to 'monitor' 'keep an eye on' or 'be aware of' the whole system:

*"You'd need the blurring of roles and you need, either you'd need someone who is on top looking down or you need a great deal of curiosity from every body involved"* [CEO of the present Italian Sportscars Factory].

Finally the integration of a design team was seen to have a substantial impact upon the success of a whole system design project. Successful integration has been observed to positively influence the other factors necessary for good whole system design; particularly 'forming and sustaining a partnership', 'human and non-human interaction', 'understanding of purpose and process' and the 'alignment of interests'. Additionally, developing an integrated team is significantly assisted by the role of the facilitatorteam leader as discussed in section 3.7 (Charnley & Lennon, 2011).

## 4 Discussing whole system design

Referring to the paper of Charnley & Lennon, 2011, it was highlighted that due to a lack of literature surrounding the process of a whole system design, it was not possible to develop a precise definition, this is now possible following the longitudinal observation of one case study and engagement with five additional cases.

*"Whole system design is an integrated and emergent approach to the design of more radically innovative and sustainable solutions. It encourages those involved to look at a problem as a whole; take multiple factors into account and utilise relationships between different parts of the problem as opposed to addressing one aspect at a time"*

In this paper it was highlighted a number of factors having been observed to substantially influence the success of a whole system design process and in doing so has created a framework to guide designers who will undertake such a process. Moreover, in this paper it was shown how this factors are traceable inside the most important Italian sportscars factory. This was demonstrated through citing how these factors are implemented in.

Additionally, in this Italian factory, the process of integrating multiple perspectives, needs and requirements is not without its challenges. Those involved within a whole system design process are recommended to look to research in the disciplines of collaborative, multidisciplinary and participatory design and also concurrent engineering.

This paper (as the application in the present Italina factory) refers also to the paper of Charnley & Lennon, 2011, in which the research outlines some of the key factors that inhibit integrative working such as the difficulty of maintaining a core design team (Lee, 2008) and the frequent lack of communication (Sonnenwald, 1996). Literature within these disciplines has also highlighted methods and techniques for improving successful integration such as the need for sense making activities (Klein, Moon, & Hoffman, 2006; Weick, 1995), the development of a shared understanding (Kleinsmann, 2006) and the use of extended social networks for access to relevant knowledge and expertise (Granovetter, 1973, 1983, Leenders, van Engelen, & Kratzer, 2003).

From what raised up, one of the key principles of whole system design is the identification and use of beneficial relationships and linkages between different parts of a system to ultimately optimise the whole. The paper shows that it is important for stakeholders to have an understanding of the benefits of taking this approach.

As previously said, one of the objectives of this paper was to investigate upon the factors capable of determining the successful development of a system design process, whose result is an optimized sustainable solution. The research highlighted that one of the key factors, to achieve these objectives, is to develop and maintain a design approach based on the characteristic long discussed in this paper (such as trans-disciplinary interests or capability of learning across different disciplines). However most of the modern challenging issues, for which a whole system design approach could be used, are often complex and request a wide range of different skills and know-how, in order to develop a reasonable solution (Katzenback and Smith (1993)). Even if there is only little literature on how trans-disciplinary skills could be developed and used within a concrete complex design approach, it is possible for stakeholders to refer to system thinking's literature (Senge (2006) and Katzenback and Smith (1993)) and face the challenge to apply those principles to their specific design problem, in order to develop an innovative and sustainable solution.

Moreover, there is still much discussion between disciplines surrounding what constitutes a system and how the use of systemboundaries can assist and hinder the process of design. A useful way of thinking of a system is to define it as a 'system of interest' (Checkland, 2000; Collins, Blackmore, Morris, & Watson, 2007). Collins et al., (2007) suggest that systems thinking involves being aware of systems of interest in their contexts and acknowledging what they are affected by and affect. System boundaries can change over time so what might be contextual at one time might be within a system of interest at another, therefore people make at least implicit boundary judgements about what lies within or outside of them (Collins et al., 2007). The paper identified the need for an alignment of interest to be formed between a system and its external environment or a design solution and its user/the consumer. Designers can learn from research which has prescribed methods and techniques surrounding how this alignment can be achieved. An example of this is the modelling of requirements to enable the successful communication of goals, targets which also assists the decisionmaking processes and makes them more transparent (Stechert & Franke, 2009). Additionally, literature in participatory design supports findings that a facilitator plays a principle role within an integrated design team by having the ability to oversee the relationships between systems (Brown, 2008; Wojanh, Dyke, Riley, Hensel, & Brown, 2001). As the role of a facilitator/team leader within a whole system design team is not clearly defined then this literature provides stakeholders with a valuable insight.

Finally, we can affirm that this paper has suggested that whole system design shares many attributes with other approaches to design and therefore it is important that we can learn from these disciplines.

## 5 Conclusions

This paper aimed to provide insight into the process of whole system design, to identify the factors that influenced its success and discover these factors inside of the most famous Italian sportscars factory, in order to demonstrate how the whole system deisgn methodology can take companies towards innovation. Whithin this paper has been presented a real complex whole system design project, with the purpose of making available, for all those interested, the possibility to expand the knowledge about the whole system design approach, and how this could be implemented to generate a successful holistic process, in order to develop an innovative and sustainable solution. Many of the multiple complex aspects of this specific design approach has been evaluated and explained using a concrete case of study based on the most famous Italian sportscar factory.

The successful implementation of this innovatine design approach conducted the most famous Italian sportscars factory to be one of the most important. It is important to remark that this implementation has been successful thanks to the open minded approach, which has permitted to collect, within the whole project, a vast amount of data without predetermined judgments, this has as a consequence that the data recorded was strongly objective. This resulted in the consolidation of the main factors that were thought to be common to a whole system design process and that could be accessible to as wide a community of designers as possible.

Towards the results of the study indicated, that there are multiple factors that influence the success of a whole system design process, it was centered the example of the most important Italian sportscars factory. The paper highlights these factors and uses examples in the Italian company to demonstrate best practise within whole system design. The identification of relationships between parts of a system to ultimately optimise the whole, the need for actors involved in the process to develop trans-disciplinary skills and the dynamics of a flattened hierarchy were identified as being some of the key necessities of whole system design (Charnley & Lennon, 2011).

The development of national and international partnerships across disciplines, thinking systemically, and involving stakeholders within the design process (Luck, 2007), are increasingly being recognised as necessary components of more sustainable design.